\documentclass[3p,times,procedia]{elsarticle}
\usepackage{nupha_ecrc}
\usepackage[english]{babel}
\usepackage{blindtext}
\usepackage{siunitx}

\volume{00}

\firstpage{1}

\journalname{Nuclear Physics A}

\runauth{D. Stewart}

\jid{nupha}

\jnltitlelogo{Nuclear Physics A}

\usepackage{amssymb}

 \usepackage{lineno}

\usepackage[figuresright]{rotating}

\begin{document}

\begin{frontmatter}


\title{Correlation Measurements of Charged Particles and Jets at Mid-Rapidity with Event 
    Activity at Backward-Rapidity in $\sqrt{s_\mathrm{NN}}=\SI{200}{GeV}$ $p$+Au Collisions at STAR}
\author{David Stewart (for the STAR Collaboration)}
\address{Wright Laboratory, Yale University, New Haven, Connecticut 06520}




\author{}

\hyphenation{ATLAS}

\address{}

\begin{abstract}

    Semi-inclusive charged jet spectra per trigger at STAR are presented binned
    by event activity (EA) as determined by the Beam Beam Counter (BBC) signal
    in the Au-going direction. The selected EA determination is motivated by
    correlations between the number of charged tracks in the  Time Projection Chamber (TPC) ($|\eta|<1$)
    and EA ($\eta_\mathrm{EA}\in[-5,-2]$) which are also presented.
    The jet spectra per trigger at high EA are suppressed relative to the
    spectra at low EA.  A PYTHIA investigation refutes that the suppression
    results from a trivial autocorrelation between jet kinematics and the
    acceptance of the EA and the TPC.

    
\end{abstract}


\end{frontmatter}



\section{Introduction}

The discovery of the quark-gluon plasma (QGP) is a principle success of heavy ion physics,
the investigation of which remains a primary focus in the field.
 In that search, small system
($p/d$+A) collisions were generally assumed to have insufficient energy densities to form a
QGP, and therefore used for comparison to benchmark QGP effects in A+A
collisions.  One principle way to quantify hot nuclear effects is via nuclear
modification factors ($R_\mathrm{AA}$), which are yields in A+A collisions
taken in a ratio to those in $pp$ collisions scaled by appropriate geometric
factors.  Separately, $p/d$+A collisions benchmark cold nuclear effects in the
A nucleus \cite{Salgado:2016jws}.

However, starting with the observation of a long range near-side ridge in high
multiplicity $pp$ collisions by CMS in 2010 \cite{Khachatryan:2010gv} most of
the signals indicative of flow in A+A collisions have now been observed in
small system collisions \cite{Ohlson:2017jkm,NagleZacj}. This has motivated
measuring jet spectra in small systems to check for other QGP-like signals. 

Measurements of minimum bias (MB) $R_{p/d+\mathrm{Au}}^\mathrm{jet}$ at the
Relativistic Heavy Ion Collider (RHIC) and the Large Hadron Collider (LHC) are,
as expected, consistent with unity
\cite{Adare:2015gla,ATLAS:2014cpa,Adam:2015hoa,Khachatryan:2016xdg}. However,
when binned by modeled geometric overlap into central (peripheral) collisions,
the ATLAS and PHENIX measurements report suppression (enhancement) for central
(peripheral) collisions. The modification increases with jet energy
 and ATLAS notes that it appears to be a function of the
Bjorken-$x$ of the proton ($x_p$) \cite{ATLAS:2014cpa}. 

ALICE has made two measurements of jet spectra binned by event class. The first
modified the method to classify events to address statistical difficulties in
determining the geometric factor, and found no modification of jet spectra with
central/peripheral binning \cite{Adam:2016jfp}.  The second reports a limit on
$p_\mathrm{T}$-independent out-of-jet-cone charged-energy transport which is
not consistent with the jet modification observed at ATLAS and PHENIX \cite{Acharya:2017okq}.
However, the measurements at ATLAS and PHENIX observed modification only at
higher $x_p$ values than those measured at ALICE.  This
is consistent with the jet modification trend as a function of $x_p$.

These previous measurements collectively motivate the STAR result presented in
these proceedings; namely, the first semi-inclusive jet measurements in small
system collisions at both (a) RHIC energies and (b) $x_p$ at which
ATLAS and PHENIX measurements report jet spectra modification.

\section{Correlation of Central Tracks and Triggers to High Backward-$\eta$ EA}
\label{correlations}

\begin{figure}[h]
\includegraphics[width=34pc]{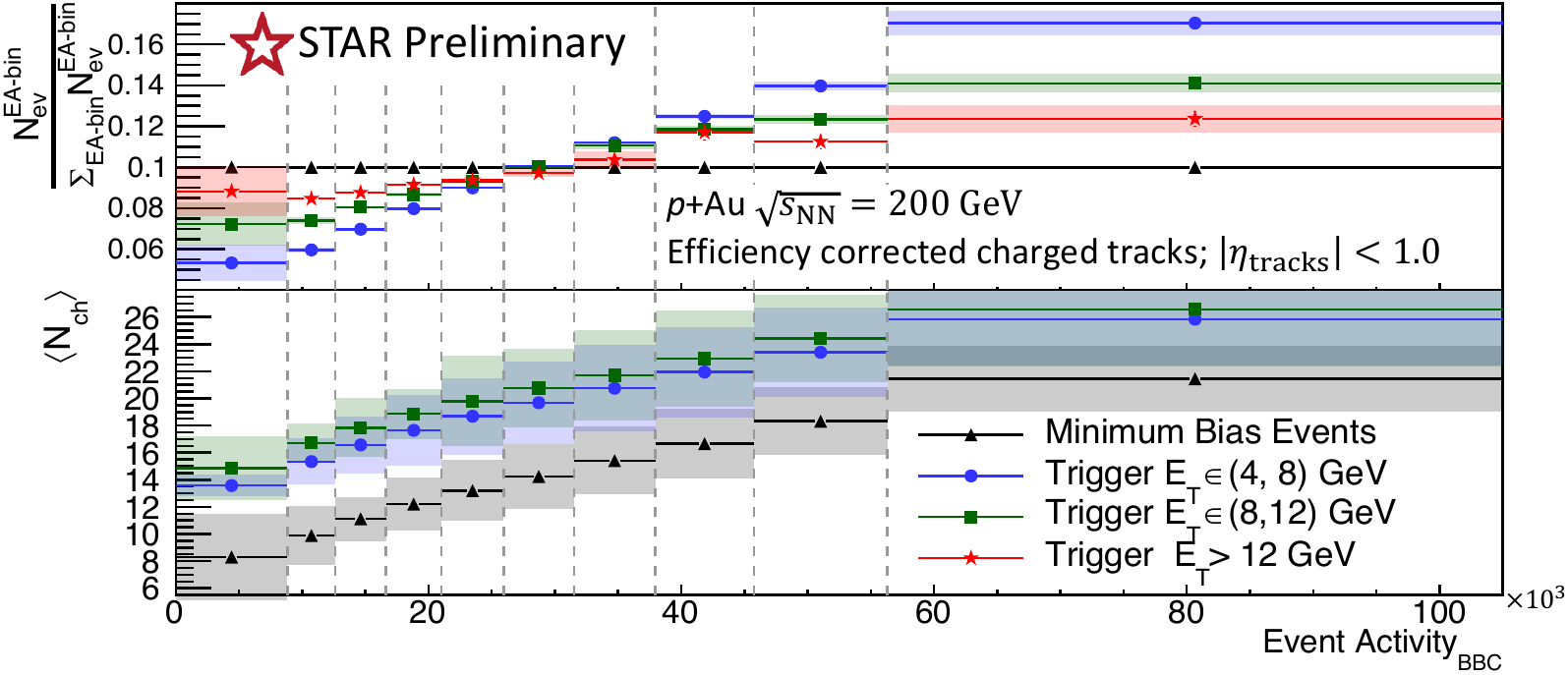}\hspace{2pc}%
    \caption{\label{fig:BBCcorr} 
    Top: Distribution of probability of events per trigger. The deciles are
    defined to contain 10\% of the MB events per bin. Bottom: Average number of
    charged tracks per event per EA decile. The values for events with Trigger
    $\mathrm{E_T}>\SI{12}{GeV}$ are statistically limited and
    therefore excluded from the results.}
\end{figure}

The event activity (EA) is defined as the sum of ADC hits in the Au-going Beam
Beam Counter (east BBC) located at $\eta\in[-5,-2]$. The EA deciles are
defined from the EA distribution in MB events.  To collect sufficient collisions with jets, a second
dataset is analyzed from events triggered by energetic hits in the Barrel
Electromagnetic Calorimeter (BEMC) which has $|\eta|<1.0$ and full azimuthal
coverage. The triggers listed in Fig.~\ref{fig:BBCcorr} are the collection of
the maximum $E_\mathrm{T}$ hit in the BEMC in each event.

The distribution of the average number of charged tracks per event
($\langle{}N_\mathrm{ch}\rangle$) per EA decile increases monotonically with
EA, as demonstrated in the bottom panel of Fig.~\ref{fig:BBCcorr}. Requiring increasing $E_\mathrm{T}$ in the
triggers, and consequently requiring higher energy jets, results in an
approximately constant addition in $\langle{}N_\mathrm{ch}\rangle$ to the MB
distribution.  This result, in addition to the rapidity gap between the EA
signal and the charged tracks, affirms the EA definition for use in binning the
semi-inclusive jet spectra.

Additionally  the
change in distribution of events with respect to EA with increasingly higher
$E_\mathrm{T}$ requirements is notable (see Fig.~\ref{fig:BBCcorr}, top panel).  A positive correlation is naively
expected and is observed.  The decrease in this positive correlation as the
trigger $E_\mathrm{T}$ requirement increases is reminiscent of a similar
observation of mid-$\eta$ charged jets and EA at high backward-$\eta$ in $p$+Pb
collisions by CMS \cite{Chatrchyan:2013gfi}, and may be a signal of theorized
physical mechanisms such as energy conservation or proton size fluctuation
\cite{McGlinchey:2016ssj}.

\section{EA binned semi-inclusive jet spectra}

\begin{figure}[h]
    \includegraphics[width=34pc]{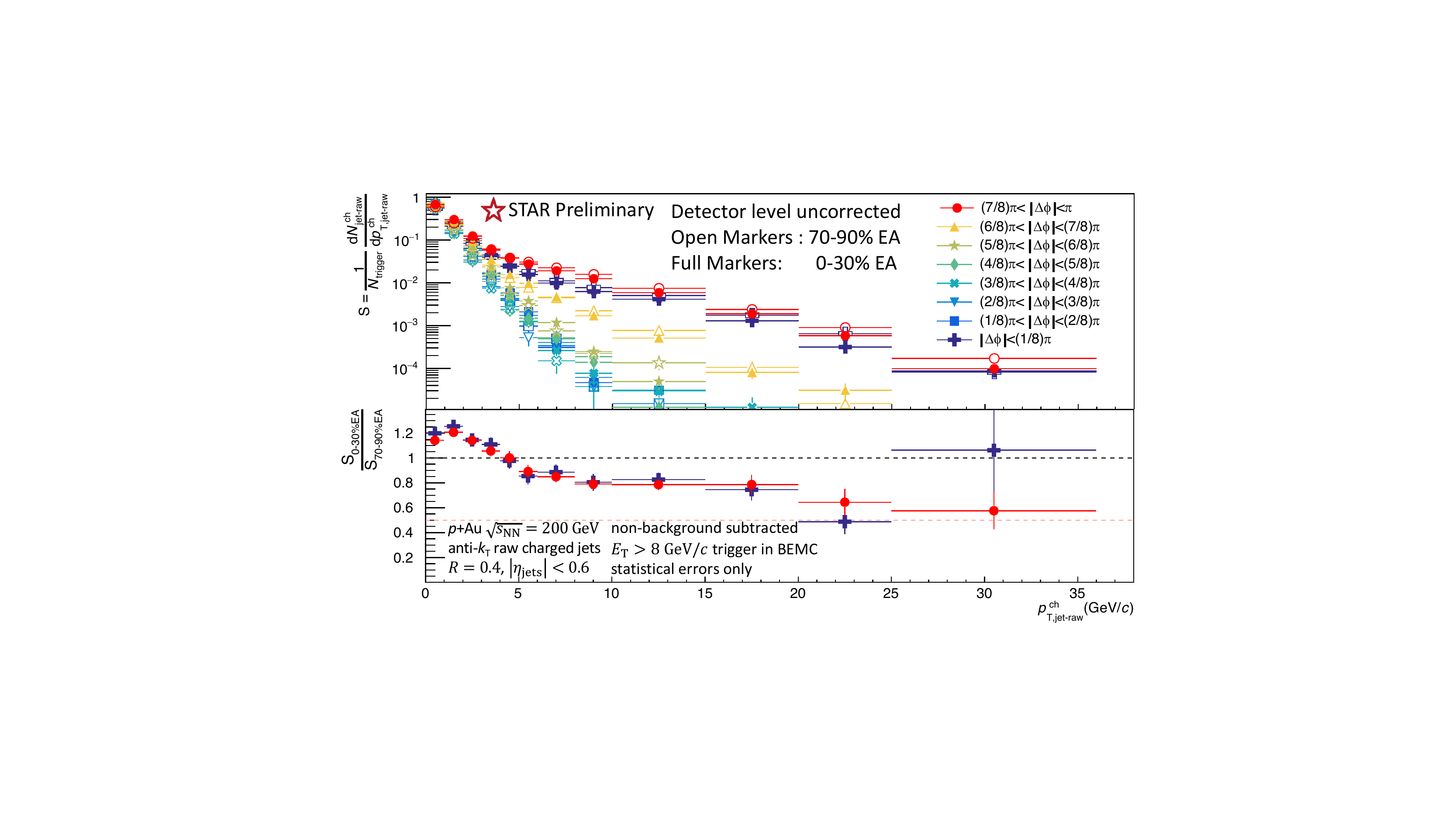}\hspace{2pc}%
    \caption{\label{fig:jet-spectra} Top: raw, uncorrected, jet spectra per
    trigger in bins of high (0-30\%) and low (70-90\%) EA, sub-divided into
    bins of $|\phi_\mathrm{jet}-\phi_\mathrm{trig}|$. Bottom: ratios of
high-to-low EA jet spectra.} \end{figure}

Events are selected with a BEMC trigger ($E_\mathrm{T-max,BEMC}>\SI{8}{GeV}$) and 
grouped into high and low EA. The high (low) EA groups correspond to the 
highest 30\% (70-90\%) of the distribution and labeled as 0-30\% (70-90\%) EA.
Within each event, charged tracks are clustered by the anti-$k_\mathrm{T}$
algorithm  \cite{Cacciari:2008gp} with $R=0.4$, and binned in azimuth relative to the trigger 
($\Delta\phi$).  The resulting jet spectra per trigger, along with high-to-low EA
spectra ratios, are shown in Fig.~\ref{fig:jet-spectra}.

The jet spectra per trigger are preliminary and uncorrected for detector
effects.  However, the spectra ratios shown in the bottom panel are not
anticipated to change within uncertainties because (a) it has been shown via
detector simulations \cite{Tongs-poster} that the charged track reconstruction
efficiency is not EA dependent and (b) there is negligible background and,
consequently, negligible combinatorial jets.  The second point is demonstrated
by  spectra in Fig~\ref{fig:jet-spectra}.  At
``jet-like'' energies,
$p_\mathrm{T,jet-raw}^\mathrm{ch}>\SI{10}{GeV/\mathit{c}}$, the transverse 
($\pi/8<|\Delta\phi|<5\pi/8$)
jet
spectra is two orders of magnitude smaller than those of trigger-side
($|\Delta\phi|<\pi/8$) and recoil-side ($7\pi/8<|\Delta\phi|$) jets.


The primary feature of the data is a marked suppression in high to low EA in
both the trigger-side and recoil-side spectra. It is of interest that both these
suppression ratios are comparable.  This is qualitatively different from jet
suppression in A+A collisions, where the recoil jets traverse more QGP on
average and are more suppressed than those on the trigger-side
\cite{Adams:2003im}.

The lower values of the trigger-side relative to the recoil-side spectra in the top
panel in Fig.~\ref{fig:jet-spectra} may be understood as a trigger selection
bias. To leading order, the trigger-side and recoil-side jets represent recoiling
jet pairs, which must balance in the sum of charged and neutral $p_\mathrm{T}$.
The trigger typically selects the jet with the leading $p^\mathrm{neut}_\mathrm{T}$,
and thus biases the near-side spectra to be higher (lower) for neutral
(charged) jets relative to the recoil-side. The data hint that, as expected,
this bias decreases at higher $p_\mathrm{T,jet}^\mathrm{ch}$ energies.



\begin{figure}[t]
\includegraphics[width=34pc]{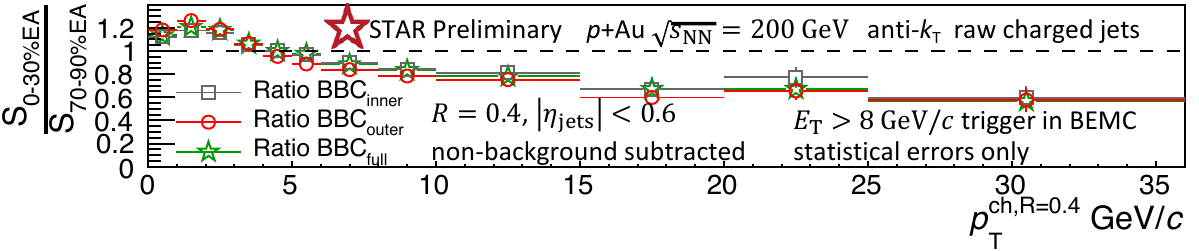}\hspace{2pc}%
    \caption{\label{fig:BBCInOut} 
    High to low EA ratio of recoil-side jet spectra ($|\phi_\mathrm{jet}-\phi_\mathrm{trigger}|>7\pi/8$) per trigger 
    when EA is determined by the inner, outer, and full BBC (the last of which is identical to values
    in bottom panel of Fig.~\ref{fig:jet-spectra}).}
\end{figure}

A PYTHIA~8 \cite{Sjostrand:2014zea} study was conducted to investigate if the
spectra suppression might result from a trivial autocorrelation in which some
jets that fall outside of the acceptance of the TPC (and therefore lower the
spectra) also hit the BBC (and therefore simultaneously
raise EA values).  PYTHIA is used to generate inclusive \SI{200}{GeV} $pp$
events. The EA signal is taken as the sum of charged particles in
$\eta_\mathrm{EA-inner}$ ($[-5,-3.4]$), $\eta_\mathrm{EA-outer}$ ($[-3.4,-2]$), and
$\eta_\mathrm{EA-inner}+\eta_\mathrm{EA-outer}$ ($\eta_\mathrm{EA}$).  Neutral
final state particles within $\eta_\mathrm{BEMC}$ are used as triggers, and
charged jets with $R=0.4$ are formed from charged particles in the TPC
acceptance.  The resulting semi-inclusive jet spectra unexpectedly show
suppression in the ratio of high to low EA-binned data. This is true for all EA
acceptances ($\eta_\mathrm{EA-inner}$, $\eta_\mathrm{EA-outer}$, and
$\eta_\mathrm{EA}$).  In order to quantify how much of this suppression results
from the proposed autocorrelation, the simulation is run a second time in which
the EA acceptance in each event is set to the BBC farthest from the two highest
$p_{T}$ jets.  For example, if the leading full jets had axes at $\eta=0.4$ and
$-0.9$, then EA acceptance (inner/outer/full) would be at the ``opposite BBC''
at $[3.4,5]$/$[2,3.4]$/$[2,5]$.  The result of this second run is that the
suppression decreases when using $\eta_\mathrm{EA-outer}$ but only minimally
when using $\eta_\mathrm{EA-inner}$.  Accordingly, the events were re-binned
using $\eta_\mathrm{EA-inner}$ and $\eta_\mathrm{EA-outer}$. The resulting
ratios of the recoil spectra are shown in Fig.~\ref{fig:BBCInOut}. As observed,
the effect is not statistically significant.  Therefore, the spectra
suppression is not a result of a trivial autocorrelation from jet kinematics
and acceptance of the TPC and EA.

\section{Conclusions and remarks}

Semi-inclusive jet spectra in both trigger and recoil azimuths are significantly
suppressed in high EA relative to low EA. This result is not yet corrected
for detector and jet reconstruction efficiency; however, the corrections
are anticipated to largely cancel in ratio such that the suppression remains.
PYTHIA simulations verify the suppression is not a trivial autocorrelation
between jet kinematics and detector acceptance. PYTHIA also unexpectedly 
predicts significant suppression. Curiously, ALICE presented a similar result
in this conference for a second jet quenching observable acoplanarity in 
$pp$ collisions \cite{Jacobs:2020ptj}. ALICE reported
an unexpectedly broadening in acoplanarity in high multiplicity events
relative to MB events accompanied by a similar currently unexplained broadening in the PYTHIA simulation.
Measuring acoplanarity modification is principally a matter of 
separating spectra into fine bins in azimuth, and a natural expansion of this STAR analysis. 







\bibliographystyle{hrev2}
\bibliography{inspire}







\end{document}